\documentclass[runningheads]{llncs}
\usepackage[T1]{fontenc}
\usepackage{graphicx}
\usepackage{algorithm}
\usepackage{algorithmicx}
\usepackage{algpseudocode}
\usepackage{hyperref}
\usepackage{nicefrac}
\usepackage{xfrac}
\usepackage{tabularx,booktabs}
\usepackage{amsmath}
\begin{document}
\title{Driver Locations Harvesting Attack on pRide}


\def\ANON{0}

\ifnum\ANON=0
\author{Shyam Murthy\orcidID{0000-0002-0222-322X} \\
Srinivas Vivek\orcidID{0000-0002-8426-0859}}
\authorrunning{S. Murthy and S. Vivek}
\institute{International Institute of Information Technology Bangalore, Bengaluru, India. \\
\email{shyam.sm@iiitb.ac.in}\\
\email{srinivas.vivek@iiitb.ac.in}}
\else
\author{ }
\authorrunning{ }
\institute{ }
\email{ }
\fi
\maketitle              % typeset the header of the contribution
\ifnum\ANON=0
\let\thefootnote\relax\footnotetext{This version of the contribution has been accepted for publication, after peer review (when applicable) but is not the Version of Record and does not reflect post-acceptance improvements, or any corrections. The Version of Record is available online at: http://dx.doi.org/10.1007/978-3-031-23020-2\_36. Use of this Accepted Version is subject to the publisher’s Accepted Manuscript terms of use https://www.springernature.com/gp/open-research/policies/accepted-manuscript-terms\\Corresponding Author: Shyam Murthy, shyam.sm@iiitb.ac.in}
%\footnote{S. Murthy and S. Vivek are with the International Institute of Information Technology Bangalore, India. Corresponding Author: Shyam Murthy, shyam.sm@iiitb.ac.in}
\fi
\def\etal{\textit{et al. }}
\begin{abstract}
% Abstract : 150--250 words
Privacy preservation in Ride-Hailing Services (RHS) is intended to
protect privacy of drivers and riders.
pRide, published in IEEE Trans. Vehicular Technology 2021, is a prediction based privacy-preserving 
RHS protocol to match riders with an optimum driver.
In the protocol, the Service Provider (SP) homomorphically computes Euclidean distances between
encrypted locations of drivers and rider.  Rider selects an optimum driver
using decrypted distances augmented by a new-ride-emergence prediction.
To improve the effectiveness of driver selection, the paper
proposes an enhanced version where each driver gives encrypted distances
to each corner of her grid.
To thwart a rider from using these distances to launch an inference attack,
the SP blinds these distances before sharing them with the rider.

In this work, we propose a passive attack where an honest-but-curious adversary rider 
who makes a single ride request and receives the blinded distances from SP can recover 
the constants used to blind the distances.
%Using the unblinded distances, the adversary obtains four equiprobable driver locations
%in each of the possible driver grids.
%Finally, the adversary uses the rider to driver distance and
%Google Nearest Road API to obtain the precise locations of responding drivers.
Using the unblinded distances, rider to driver distance and Google Nearest Road API,
the adversary can obtain the precise locations of responding drivers.
We conduct experiments with random on-road driver locations for four different
cities. Our experiments show that we can determine the precise locations of
at least 80\% of  the drivers participating in the enhanced pRide protocol.

\end{abstract}

\section{Introduction}
\label{tvt_attack:intro}
According to a recent research by MordorIntelligence \cite{mordorIntel},
the global Ride-Hailing Services (RHS) market, valued at 
USD 113 billion in 2020, is expected to reach USD 230 billion by 2026.
With such a huge reach, individual privacy and security issues
are always of primary concern.  Ride-Hailing Service Providers (SP) 
like Uber, Lyft, Ola provide services in many parts of the world.
Among other features, the SP facilitates ride booking and fare payment options 
for their customers, namely riders who subscribe with the SP for RHS.
Drivers of vehicles such as cars and motorcycles sign-up with the SP
in order to offer rides.
At the time of subscription, the SP collects private information of riders and
drivers in order to provide services effectively as well as required
by local governance laws.
In addition, the SP collects statistics of riders and drivers
for every ride that is offered in its network.
This naturally brings up the topic of individual data privacy concerns
from both riders as well as drivers over their data held by the SP.
Also, curious or malicious drivers or riders might be interested in
learning more about the other parties.
There are a number of works and their analysis in the literature that look at 
privacy-preserving RHS, we list some of them in Section \ref{sec:rel_works}.

Huang \etal proposed \textit{pRide} \cite{pRideHuang2021}, a privacy-preserving 
online RHS protocol that aims to provide the optimum driver in a global 
perspective thereby minimizing the unnecessary travel distance to pick 
the rider.  
The protocol makes use of a deep learning model to predict emergence
of new ride requests in a ride-hailing region to enable the SP to 
make use of such prediction while matching optimum drivers to ride requests.
They show that by using such a prediction model in a global perspective, 
the overall distance travelled by a matching driver is minimized
compared with matching a nearest driver in the local region.
The protocol proposes to use a Somewhat Homomorphic Encryption (SHE)
scheme to encrypt rider and driver locations.
The advantage of using a homomorphic encryption scheme is that it allows
computations on ciphertexts so that the result of computation is available
only after decryption.   Fully Homomorphic Encryption (FHE) schemes 
that support potentially any number of homomorphic operations have 
high cost in terms of large ciphertexts and high computation latency.
Hence, many practical applications that know, a priori, the bound on the number
of homomorphic operations, prefer to use SHE schemes.
In the pRide paper, the authors use the FV cryptosystem \cite{EPRINT:FanVer12}
in the implementation of their scheme.
Even though applications make use of semantically secure cryptosystems,
careful analysis is required to make sure no unintended security
holes are introduced while adapting the cryptosystem to their applications.

The pRide protocol, described in more detail in Section \ref{sec:pride},
has two parts, the basic protocol and an enhanced version.
We discuss the basic protocol in this paragraph.
In the initialization phase, SP divides the area of its operation into grids,
the details of which are made available to all parties.  
SP keeps a record of ride requests emanating from each grid over specific time 
epochs and trains a prediction model using this information.  It 
then uses this information to predict the grid-based distribution of requests 
for the next period, denoted by $PR(g)$, namely the prediction result for grid id $g$.
Drivers, registered with the SP, submit their current
grid id to the SP so that the SP can maintain the driver distribution map.
A rider who wishes to hail a ride, picks a (public key, secret key) pair, 
encrypts her coordinates and sends the ciphertext and public key to SP 
along with the ride request.  When SP receives the ride request, 
it performs a search for a suitable driver 
in a preset order of grids around the rider's grid 
and obtains a list of candidate drivers using the driver distribution map.
SP then forwards the ride request to 
all candidate drivers.  To offer their ride, drivers respond to SP by encrypting their location using
the rider's public key.  SP then homomorphically computes the square of the Euclidean
distance between rider and drivers' encrypted locations and forwards
the same to the rider along with $PR(g)$ where $g$ is the driver's grid id.
Rider decrypts the distances\footnote{Henceforth in the paper, we use the term distance
to mean squared Euclidean distance.} and picks the shortest distance $D_0$.  It then
performs two checks over the list of sorted distances.  First, is $D_i - D_0 < D_0 - D_{diag}$$?$, where $D_i$ is
the distance for $i\textsuperscript{th}$ driver and $D_{diag}$ is the length of the
diagonal of the grid, and second, does the model predict \textit{no} new ride request
emerging in the driver's grid within a short time period$?$  When both these conditions are
satisfied, the rider informs SP about the selected index $i$ after which
the SP facilitates secure ride establishment with the rider and selected driver.

In order to optimize their ride matching, the paper proposes
\textit{enhanced pRide}  built on top of the basic pRide protocol,
but having a different method to pick the optimum driver.
They show that they get better results when a driver 
\textit{also} provides her encrypted distance to the farthest corner of her grid. 
This way the rider can use that distance, instead of $D_{diag}$ in the 
aforementioned check to select the optimum driver.  
However, the authors notice that if such
a distance is decrypted by an adversarial rider, she can launch
an inference attack to obtain driver's locations.  In order to thwart such
an attack, the paper proposes a novel method where the driver provides SP
with her encrypted distances to the four corners of her grid.  SP then
picks random integers to homomorphically blind the distances before sharing
the same with the rider.  Rider then decrypts the blinded distances
and applies a \textit{private comparison algorithm} which determines the result
of the inequality $D_i - D_0 < D_0 - D_{maxdist}$, where $D_{maxdist}$
is the distance between the driver and the farthest corner of her grid $g$.
Finally, using this inequality and $PR(g)$, it outputs 
the optimum selected driver.

As described earlier, in the enhanced pRide protocol, the SP homomorphically 
blinds the encrypted distances 
with random integers before sharing them with the rider.  
In this paper, we show that such a blinding scheme is insecure, whence 
an adversary rider can recover the underlying distances and then 
deduce the locations of at least 80\% of the drivers responding to a 
single ride request of the rider when using the enhanced pRide protocol.

\subsection{Comparison with ORide \cite{ORidePaper} Protocol}
\label{tvt_attack:oride_ref}
The pRide paper shows that their enhanced scheme is more effective
with the same level of security as that of the basic version with only a 
small compromise in its efficiency.  In addition, by way of experiments, 
they show their computation cost is significantly better compared to a 
state-of-the-art protocol named ORide \cite{ORidePaper}.
We note here that the method in the basic pRide protocol where the SP employs 
the homomorphic property of SHE to compute the Euclidean distance between 
driver and rider to share the encrypted distances with rider is identical 
to what is described in the ORide paper.  The part that is different
is that in the ORide paper to pick the nearest driver, only drivers inside the rider's grid are
chosen as candidate drivers, whereas in the pRide protocol, only drivers outside
the rider's grid are candidate drivers so as to optimize in a global perspective.

In \cite{deepakOridePaper}, Kumaraswamy \etal demonstrated a driver locations
harvesting attack by honest-but-curious riders on the ORide protocol, where they 
determine the exact locations of 40\% of drivers participating in the ORide protocol.
In the same paper, the authors also provide a mitigation solution 
wherein a driver gives her perturbed location instead of her actual location.
The aforementioned attack on the ORide protocol and the mitigating solution
are both applicable to the basic pRide protocol.

In \cite{murthyOrideAttack}, Murthy \etal demonstrated a driver locations harvesting
attack, again by honest-but-curious adversary riders, using triangulation on the 
ORide protocol, where they show that they can determine the exact locations of all 
participating drivers in the ORide protocol.
Further, they extend their method onto the mitigation solution suggested
by \cite{deepakOridePaper} and show that they can determine locations of between 25\%
to 50\% of the participating drivers.

As mentioned earlier, in the pRide protocol, the method where the rider obtains encrypted driver distances 
is identical to that in the ORide protocol.  Due to this, any
location harvesting attack on ORide, like in the cases of \cite{deepakOridePaper}
and \cite{murthyOrideAttack},
are also directly applicable to the basic pRide protocol.
%In Section \ref{sec:attack_mitigation}, we examine the effectiveness of our attack 
%when the mitigation solution provided by \cite{deepakOridePaper} 
%is adapted onto the pRide protocol.

\subsection{Our Contribution}
We present a passive driver location harvesting attack on the {\em enhanced} pRide protocol.  
The honest-but-curious adversary rider issues a single ride request with
a search radius ($SR$) = 1, such that grids adjacent to the rider's grid 
are searched (as explained in the pRide paper, Section V-B-4, pp. 6).
In our attack, the adversary rider receives, per driver, a set of 
encrypted blinded distances between the driver's location and each corner of the driver's grid.  
One would expect that such a blinding process would make it hard for 
the rider to deduce anything about the underlying distances.

Rider decrypts the ciphertexts received from SP to obtain blinded distances.
Next, by computing the Greatest Common Divisor (GCD) of the blinded distances and eliminating
common factors, the rider recovers the
blinding values after which the distances are easily obtained.
Rider now has the four distances from driver to each corner of the driver's grid.
Using these distances, the rider computes four equiprobable 
driver locations in each of the four grids adjacent to the rider's grid.
This is due to the fact that the distances are in random order and, so, there is 
no correlation between each corner of the grid and its distance to the driver.
Rider knows the distance between herself and each responding driver.
Now, using the distance between herself and a particular
responding driver (say, $\delta$), 
the rider draws a rider-circle with center as her location and radius 
$= \delta$.  Probable driver locations that lie on the rider-circle
are filtered in and in case multiple such locations are obtained,
Google Nearest Roads API \cite{googleroadapi} is used to output one 
location that is closest to a motorable road.
We conduct our experiments using rectangular grids on four different cities
around the world and the results are summarized in Table \ref{pride:tab:results}.  
We show that we can obtain exact driver locations of up to
80\% of drivers who respond to a rider's request.

Our attack invalidates Theorem 4, pp. 9, of the pRide paper \cite{pRideHuang2021}, 
which states that pRide is adaptively $\mathcal{L}_{access}$ semantically secure 
against semi-honest adversaries, where $\mathcal{L}_{access}$ gives the access pattern
of the SP and rider, which is simply the list of drivers that respond
to a specific ride request.
Hence, when our attack is combined with that in \cite{murthyOrideAttack}, the driver 
location security of the pRide paper is fully compromised, and so is the mitigation
solution of \cite{deepakOridePaper} if applied to the basic pRide protocol.
We stress that the attack from \cite{murthyOrideAttack} is not directly applicable to 
the pRide protocol, but works only in combination with our attack.

The rest of the paper is organized as follows.   Section \ref{sec:pride}
describes the pRide protocol.  Section \ref{sec:our_attack}
describes our attack.  Section \ref{sec:exp_res} gives details about our experiments
and results.  Section \ref{sec:rel_works} gives some of the recent works in 
privacy-preserving RHS, followed by conclusions.

%
%\section{Overview of pRide \cite{pRideHuang2021}}
\section{Overview of pRide Protocol}
\label{sec:pride}
In this section, we provide an overview of the pRide protocol followed by
a description of the threat model adopted therein.
For more details, the interested reader is referred to the original paper
\cite{pRideHuang2021}.\\
\noindent\textit{Remark}: Unless qualified as enhanced or basic, we will use the 
term \textit{pRide protocol} to refer to the complete pRide protocol, 
consisting of both the basic and enhanced parts.

\subsection{pRide Protocol}
\label{sec:pride_prot}
pRide is a privacy-preserving online ride-hailing protocol augmented with
a grid-based rider emergence prediction. 
The key objective of the protocol is to achieve optimum driver selection in
a global perspective instead of picking the nearest driver as done in other works \cite{ORidePaper,Pham2017PrivateRideAP}.
Selecting such a driver might be a better choice in order to minimize the overall
empty travel distance traversed by drivers to pick up riders in the whole system.
The prediction of requests based on deep learning plays an important role in driver selection.

The protocol has two parts, the basic protocol and an enhancement, built on top of the basic protocol, are summarized in the following steps.  Steps \ref{item:basic_one} to
\ref{item:basic_last} constitute the basic pRide protocol, followed by steps of the enhanced pRide protocol.
\begin{enumerate}%[\label=\textbullet]
\item \label{item:basic_one} The three parties involved in the pRide protocol are: driver, 
rider and service provider (SP).
The SP does not collude with either rider or drivers.  The SP as well as the users, namely, drivers and riders, are honest-but-curious entities 
who execute the protocol correctly, but are keen on
knowing about each other's sensitive information.
The protocol aims to protect all users' privacy from other riders and drivers, 
such that the precise location of one party is not learnt by the other party
during the ride matching process.  However, only after a driver is matched with
a rider, they start to communicate through a secure channel.
\item During system initialization, the SP divides its area of its operation into
rectangular grids of suitable sizes (size is based on sufficient ride density
so as to maintain rider anonymity) and publishes the same.
For example, a city like New York City together with its surrounding boroughs, where 
the SP is allowed to provide rides as permitted by local authorities, can be
termed as the SP's area of operation.
\item Drivers, available to offer rides, submit their real-time grid id to
the SP to enable it to maintain a driver distribution map.
\item Rider, wishing to hail a ride, generates a key pair (public key $p_k$, private key $s_k$)
from the FV SHE scheme \cite{EPRINT:FanVer12}, encrypts her location using
$p_k$, and submits a ride-request along with her location ciphertext, 
her current grid id and $p_k$ to the SP.
The FV SHE scheme works on integers, hence, the coordinates of users are 
encoded as integers using UTM format\footnote{Universal 
Transverse Mercator: a map-projection system for geographical locations \cite{UTMGrid}.}.
\item SP keeps a record of ride requests in each grid and maintains a 
real-time ride request distribution map in every time period.  
It makes use of Convolutional long short-term memory (Convolutional LSTM \cite{shi_clstm})
to train a prediction model with the ride request distribution information.
Based on a temporal sequence of grid information, SP obtains prediction
result $PR(g)$, a non-negative integer which predicts the number of requests 
in the next time period for grid id $g$.
\item As soon as SP receives the ride request, it performs a driver search   
with a search radius ($SR$) in a preset order of grids starting with the grid nearest
to rider.  The rider's grid is not searched so as to avoid the nearest
driver who would always be found in the rider's grid.
When $SR=1$, only grids adjacent to the rider are searched.
Using the driver distribution map, SP creates a list of candidate drivers
and forwards the ride-request to all such drivers.
\item When the $i\textsuperscript{th}$ driver $d_i$ receives the ride-request, she
encrypts her location using $p_k$ and forwards it to SP.
\item SP homomorphically computes the square of the Euclidean distance between
the rider and drivers' locations.  It then forwards these distances to rider
along with driver id $i$ and $PR(g_i)$, $g_i$ is $i$'s grid id.
\item \label{ridesel_check} Rider uses $s_k$ to decrypt the distances and sorts them to obtain the
smallest distance $D_0$.  For each distance in the sorted list, she runs the 
following two checks to pick the optimum driver:
\begin{enumerate}
\item $2 D_0 - D_i > D_{diag}$, where $D_i$ is
the distance for $i\textsuperscript{th}$ driver and $D_{diag}$ is the length of the
diagonal of the grid.
\item $PR(g_i)$, where $g_i$ is the driver's grid id, which checks if no new ride 
request is emerging in a short time in grid $g_i$.
\end{enumerate}
\item \label{item:basic_last} As soon as both the aforementioned conditions are 
satisfied, rider determines the optimum driver and informs the same to SP to 
continue with secure ride establishment between rider and selected driver.
\item In order to improve the effectiveness of driver selection, the authors
notice that they can minimize the empty distance travelled by the driver
by using $D_{maxdist}$ instead of $D_{diag}$ in the ride selection check (Step \ref{ridesel_check}),
where $D_{maxdist}$ is the distance between the
driver and the farthest corner in her grid.
However, the authors realize that an adversary rider, after decryption,
can use $D_{maxdist}$ to launch an inference attack to obtain driver's 
precise location.  
They, therefore, propose \textit{enhanced pRide} to thwart such an attack.
\item In the enhanced pRide protocol, each driver, in addition to sending
encryptions of her coordinates, also sends the encryptions of distances to
each corner of her grid to the SP.
\item To pick the optimum driver, rider now needs to perform the check 
$2 D_0 - D_i > D_{maxdist}$, for each driver $i$,
using a \textit{private comparison algorithm}, as explained below (Steps \ref{item:one}, \ref{item:SP} and \ref{item:two}).
\item \label{item:zero} As in the earlier basic pRide protocol, rider receives a list of distances
to each of the candidate drivers, decrypts them and selects the smallest $D_0$.
\item \label{item:one} In order to find the optimum driver, for each $D_i$, $i > 0$, rider sets $D' = 2 D_0 - D_i$, encrypts $D'$ as  
$\widetilde{D'}$ and sends $\widetilde{D'}$ and $i$ to SP.  
\item \label{item:SP} SP receives encrypted distances
to each of the four corners of the $i\textsuperscript{th}$ driver's grid as 
$(\widetilde{D_{ll}}, \widetilde{D_{lu}}, \widetilde{D_{rl}}, \widetilde{D_{ru}})$. 
SP generates random positive blinding integers $e$ and $r$, and homomorphically blinds
each of the ciphertexts as
\begingroup\makeatletter\def\f@size{10}\check@mathfonts
\begin{equation}
\begin{split}
	\widetilde{V'} & = e\cdot\widetilde{D'} + \widetilde{r}  \\
	\widetilde{V_{ll}} & = e\cdot\widetilde{D_{ll}} + \widetilde{r}  \\
	\widetilde{V_{lu}} & = e\cdot\widetilde{D_{lu}} + \widetilde{r} \\
	\widetilde{V_{rl}} & = e\cdot\widetilde{D_{rl}} + \widetilde{r}  \\
	\widetilde{V_{ru}} & = e\cdot\widetilde{D_{ru}} + \widetilde{r}.
\end{split}
\label{pride:eqn_one}
\end{equation}
\endgroup
It then sends each of these blinded values to rider.\\
\noindent\textit{Remark}: Homomorphic addition of two ciphertexts, and homomorphic multiplication
of ciphertext with plaintext can be done very efficiently in SHE.
\item \label{item:two} Rider decrypts each of these blinded values and compares $V'$ with each of
$(V_{ll}, V_{lu}, V_{rl}, V_{ru})$.  If $V'$ is greater than all the four
values, then it implies that $D' > D_{maxdist}$.
\item \label{item:three} Rider then uses this comparison result and $PR(g)$ value as in the
basic pRide protocol to select the optimum driver and informs the same to SP.
If these checks fail, then the Steps \ref{item:one} through \ref{item:three} are repeated until an optimum 
driver is obtained by walking through each entry in the candidate driver list.
\end{enumerate}

The authors evaluate the performance of their enhanced pRide protocol over
real-world datasets.  Their results show that their protocol is effective
in saving empty distance as well as in maintaining drivers' privacy during
the ride matching process.
Finally, they compare the basic and enhanced versions of pRide
and prove that the latter is more effective in choosing the optimum driver
with the same level of privacy.
The security of their protocol is based on the apparent hardness of 
retrieving the blinding parameters when given only the blinded values.

In our attack described in Section \ref{sec:our_attack}, we show that we can determine the underlying
distance values when given only their blinded values, where 
blinding is done as described in Step \ref{item:SP}.
We then go on to use the distances to get the precise coordinates
of responding drivers.

\subsection{Threat Model}
We consider the same threat model considered in the pRide protocol, where all parties, namely the SP, drivers and riders, are
honest in executing the protocol.  Riders submit valid requests by encrypting
their correct coordinates to the SP, and the drivers also submit the
encryptions of their current coordinates to the SP.
SP does not collude with either drivers or riders.  Drivers do not
collude with riders.

All parties are honest-but-curious in the protocol.  Thus, each party is curious
to know more about the sensitive information of the other party.
In particular, riders are curious to know about drivers' locations and vice-versa.
pRide also considers the case of an adversary rider who follows the
protocol correctly but launches an inference attack by performing private computations
on received driver coordinates to infer drivers' precise locations, and
so the authors propose enhanced pRide to thwart such an attack.
Their paper aims to preserve driver and rider location information from SP, and
to preserve driver location information from rider.

In this paper, we consider the same threat model to model the adversaries.
The ride request issued by an honest-but-curious adversary rider is
indistinguishable from a ride request issued by any other legitimate rider
in the protocol.
In a real-life scenario, a competitor SP with the intention of harvesting
driver information of another SP, can mount such an attack without 
being detected by the target SP.
%We look at driver location-harvesting attack by adversary rider on drivers
%participating in the enhanced pRide protocol.
%The adversary rider is honest-but-curious and follows the protocol steps correctly.
%In order to mount the attack, the adversary rider issues a single ride request
%which is indistinguishable from a ride request issued by any other legitimate 
%rider in the protocol.
%The adversary receives responses as per the protocol but uses the information 
%to perform driver location-harvesting attack.

%
\section{Our Attack}
\label{sec:our_attack}
In this section, we present our driver location harvesting attack on
the enhanced pRide protocol by a honest-but-curious adversary rider ($R$).
$R$ issues a single ride request as per the pRide protocol.
SP will not be able to distinguish between a ride request issued by an
adversary rider versus another by a legitimate rider.
In this section, for ease of exposition, we explain the recovery of location 
of one particular driver $D_p$, who has responded to ride request by
$R$, shown in Figure \ref{fig:tvt_attack}. 
$D_p$ is located at distance $\delta$ from $R$.
Our attack extends easily to all responding drivers, since each response is
handled independently by the SP.

\subsection{Retrieving Distances}
\label{tvt:retr_dist}
$R$ issues a ride request as per the pRide protocol with
search radius $SR = 1$.  By this, only the grids adjacent to the rider's grid
are searched by SP for candidate drivers.

We recall here the steps of pRide and enhanced pRide protocols from Section \ref{sec:pride_prot}. 
In Step \ref{item:zero}, the rider $R$ obtains the distances between
herself and all the responding drivers in the clear (distance between $R$ and $D_p$ is $\delta$).
In addition, from Step \ref{item:SP}, $R$ receives the ciphertexts
$(\widetilde{V'}, \widetilde{V_{ll}}, \widetilde{V_{lu}}, \widetilde{V_{rl}}, \allowbreak \widetilde{V_{ru}})$,
which after decryption gives $(V', V_{ll}, V_{lu}, V_{rl}, V_{ru})$.
We know that $\widetilde{D'}$ is the encryption of $2 D_0 - \delta$, and 
\begingroup\makeatletter\def\f@size{10}\check@mathfonts
\begin{equation}
\begin{split}
	V' = \, & e\cdot\widetilde{D'} + \widetilde{r} \\
	V_{ll} = \, & e\cdot\widetilde{D_{ll}} + \widetilde{r} \\
	V_{lu} = \, & e\cdot\widetilde{D_{lu}} + \widetilde{r} \\
	V_{rl} = \, & e\cdot\widetilde{D_{rl}} + \widetilde{r}  \\
	V_{ru} = \, & e\cdot\widetilde{D_{ru}} + \widetilde{r},
\end{split}
\end{equation}
\endgroup
where $e$ and $r$ are the blinding integers chosen by SP. 

$R$ then computes the difference of every pair from $(V_{ll}, V_{lu}, V_{rl},\allowbreak  V_{ru})$, 
decrypts them using her secret key and stores them as $(P, Q, R, S, T, \allowbreak U)$, in no particular order. \\
%\textit{Remark}: An unrelated work on integer polynomial recovery \cite{MurthyV19},
%when given only the outputs, uses factors of differences
%of outputs to recover a monotonic polynomial.\\
The differences, thus obtained, are
\begingroup\makeatletter\def\f@size{10}\check@mathfonts
\begin{equation}
\begin{split}
	\setlength\abovedisplayskip{0pt}
	P = \, & V_{ll} - V_{lu} = e\cdot(D_{ll} - D_{lu}) \\
	Q = \, & V_{ll} - V_{rl} = e\cdot(D_{ll} - D_{rl}) \\
	R = \, & V_{ll} - V_{ru} = e\cdot(D_{ll} - D_{ru}) \\
	S = \, & V_{lu} - V_{rl} = e\cdot(D_{lu} - D_{rl}) \\
	T = \, & V_{lu} - V_{ru} = e\cdot(D_{lu} - D_{ru})  \\
	U = \, & V_{rl} - V_{ru} = e\cdot(D_{rl} - D_{ru}).
\end{split}
\end{equation}
\endgroup

It can be easily seen that the GCD of any two of $(P,Q,R,S,T)$, say $P$ and $Q$,
will give either $e$ or its multiple.
The latter case will occur when $ (D_{ll} - D_{lu})$ and $ (D_{ll} - D_{rl})$ are not relatively prime, and by eliminating any common factors between them,
we can hope to retrieve the exact value of $e$ with a high probability. \\
\textit{Remark}: The probability of $n$ randomly chosen integers being
coprime is $\frac{1}{\zeta(n)}$, where $\zeta$ is the Riemann Zeta function  \cite{coprime_wiki}, and for two such integers the probability is $\frac{6}{\pi^2}$.
This means in about 60\% of cases we can find the value of $e$ straightaway, and in
rest of the cases we can try to eliminate common factors. \\
Notice that each of the $D_{xy}$ values are squares of the Euclidean distance
between the driver's location and each corner of her grid.
Let the driver's coordinates (to be determined) be $(x,y)$ and the known corners of her grid
be $(x_1, y_1), \allowbreak (x_2, y_2),\allowbreak  (x_3, y_3)$ and $(x_4, y_4)$.
W.l.o.g, 
\begin{align}
	D_{ll} = {(x_1 - x)}^2 + {(y_1 - y)}^2 \\
	D_{lu} = {(x_2 - x)}^2 + {(y_2 - y)}^2.
\end{align}
Hence, \hspace{0.04in}
$P = e\cdot\Bigl(\bigl({(x_1 - x)}^2 + {(y_1 - y)}^2\bigl) - \bigl({(x_2 - x)}^2 + {(y_2 - y)}^2\bigr)\Bigr)$, 
which simplifies to 
\begin{eqnarray}
P = e\cdot\bigl((x_1 - x_2)(x_1 + x_2 - 2x) + (y_1 - y_2)(y_1 + y_2 -2y)\bigr).
\end{eqnarray}
By eliminating common factors, if any, we obtain
\begin{eqnarray}
	P' = \sfrac{e \cdot P}{\bigl( \textnormal{\scriptsize{GCD}}(x_1 - x_2, y_1 - y_2) \ast \textnormal{\scriptsize{GCD}}(2, x_1+x_2, y_1+y_2)\bigl)}.
\end{eqnarray}
And similarly, we get $Q', R', S', T', U'$.
Finally, $\textnormal{GCD}(P',Q',R',S', T',\allowbreak U')$ gives the value of $e$. \\
\textit{Remark}:  The coordinates of each of the grids are known at system
initialization time.  Hence, any common factors between the coordinates can be 
computed offline.

In Step \ref{item:one}, rider has the value of $\widetilde{D'}$, using which
the value of $\widetilde{r}$ is obtained from 
$V' = e\cdot\widetilde{D'} + \widetilde{r}$.
And, finally, using $e$ and $\widetilde{r}$,  
$(\widetilde{D_{ll}}, \widetilde{D_{lu}}, \widetilde{D_{rl}}, \widetilde{D_{ru}})$,
and, hence, $(D_{ll}, D_{lu}, D_{rl}, D_{ru})$ are obtained.\\
\textit{Remark}: In case we obtain a negative value for $\widetilde{r}$,
it implies that our recovery of $e$ is in error.

\subsection{Retrieving Driver Locations}
\label{tvt:retr_locs}
$R$ does not know the correlation between
the $D_{xy}$ distances and the corners of the grid as they
are distances given in random order.
In addition, since the search radius $SR=1$, any of the four grids adjacent 
to the rider's grid can be a potential grid of driver $D_p$.

\begin{figure*}[htpb]
\begin{center} % remove for 2 col
\includegraphics[width=\textwidth]{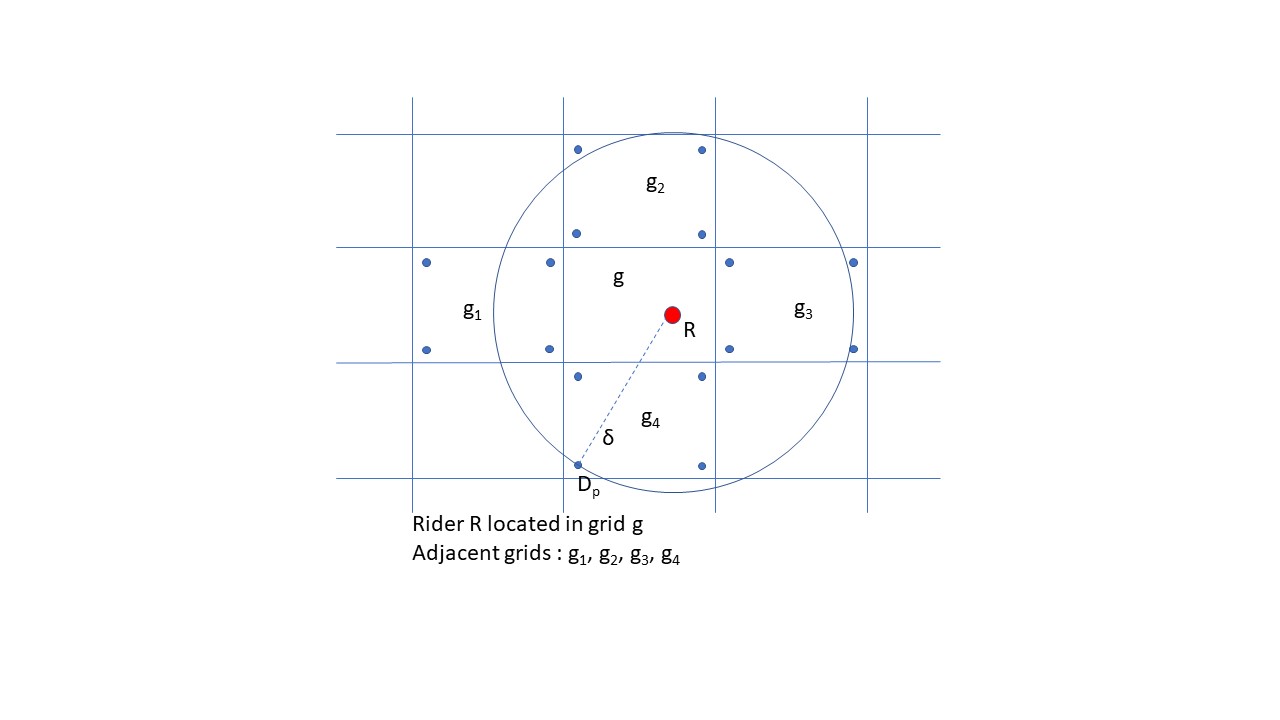}
\vspace{-.8in}
\caption{Recovered driver locations shown as small dots each in grid $g_1$ through $g_4$.} \label{fig:tvt_attack}
\end{center} % remove for 2 col
\end{figure*}

Using the four distance values $(D_{ll}, D_{lu}, D_{rl}, D_{ru})$ as radii 
and each of the respective grid corners as center of circles, 
rider obtains four points in 
each grid where all the four circles intersect.
These points, in their respective grids, represent the equiprobable locations of driver $D_p$.
Figure \ref{fig:tvt_attack} gives a pictorial view of our attack.
Adversary rider $R$ is located in grid $g$.
Driver $D_p$ is located in grid $g_4$, at a distance $\delta$ from $R$.
Each of the four probable driver locations in each adjacent grids
$g_1$ through $g_4$ are shown as small blue dots in each grid.

Using the distance between $R$ and $D_p$, namely $\delta$,
$R$ draws a rider-circle of radius $=\delta$ around herself.
As long as the driver has reported her correct coordinates, it is guaranteed that
at least one of the 16 equiprobable 
driver locations will lie on the circumference of the rider-circle.
If more than one such location is obtained, then the rider makes use of 
Google Nearest Road API \cite{googleroadapi} to find the nearest road to
each of such locations.  Since we assume that the driver is located on a motorable
road, the adversary algorithm will output the location closest to the 
nearest road.

\subsection{Analysis of our Attack}
As described in Section \ref{sec:pride_prot}, the pRide protocol makes use of a semantically 
secure cryptosystem, namely the
FV SHE scheme \cite{EPRINT:FanVer12}, to encrypt the locations of drivers and rider,
using which driver distances are computed homomorphically.
In order to pick the closest driver, the distances need to be sorted which will need
a high-depth circuit resulting in an inefficient implementation with SHE.  Hence,
the rider, in the basic pRide protocol, receives all encrypted distances, decrypts and sorts
them to pick the closest driver efficiently.  Using the distances to all drivers, 
the rider is able to perform the attacks described in 
\cite{deepakOridePaper} and \cite{murthyOrideAttack}, on the basic pRide protocol.

As described in Section of \ref{sec:pride_prot}, the enhanced pRide protocol, 
SP homomorphically blinds the distances to the four corners of drivers' grid, 
using random positive integers (Eqn. \ref{pride:eqn_one}).
However, as we show in Section \ref{tvt:retr_dist}, this blinding method is insecure.

The mitigation solution of \cite{deepakOridePaper}, where the locations are perturbed, can be applied
to the pRide protocol.  While the attack of \cite{murthyOrideAttack} is still applicable 
on the basic pRide protocol, we look at our attack on its enhanced version, when the 
mitigation solution is applied to the pRide protocol.
In that case, in response to a ride request, the driver would pick a uniform random 
location inside a circle of radius $\tau$ around her original location.
She then sends the encryption of that random location to the SP, as well as
the encrypted distances from the random location to each of the corners of her grid.
We note that $\tau$ should not be too
large, as that would have an adverse effect on driver selection by rider.
Our attack, where we retrieve the distances to grid corners, described in 
Section \ref{tvt:retr_dist}, would be applicable without
any change.  However, one of the retrieved location(s), in this case, would be the random 
location picked by the driver instead of her actual location.
The adversary could then apply the attack of \cite{murthyOrideAttack} to 
uncover the actual driver locations, with a high probability.
Since the retrieved locations might not be on a motorable road due to perturbation, 
the effectiveness of being able to use Google Nearest Road API to 
retrieve driver locations need to studied.

%\textit{Remark 2}: For completeness, we list here two trivial modifications that
%can be used to thwart the attack on basic pRide protocol.  Instead of using the
%Euclidean norm, the SP might try to use the $p$-norm between the driver and rider
%coordinates.  But as shown in \ref{deepakOridePaper}, their attack is applicable
%when the SP uses $p$-norm.

%
\section{Experiments and Results}
\label{sec:exp_res}
We use Sagemath 8.6 \cite{sagemath} to implement our attack described
in Section \ref{tvt:retr_dist} where we retrieve driver distances.
The attack, described in Section \ref{tvt:retr_locs}, where we 
retrieve the driver locations, was
implemented in Python and used the Google Nearest Road APIs for Python
\cite{googleroadapi_python}.
Both parts of the attack were executed on a commodity laptop 
with 512 GB SSD and AMD Ryzen 5 processor.
Our Sagemath and Python programs are available at: \\
\href{https://github.com/shyamsmurthy/nss2022}{https://github.com/shyamsmurthy/nss2022}.

\subsection{Experiment Details}
Our experiments were run on grids of size about 4$km\textsuperscript{2}$
superimposed on maps of 4 large cities around the world, namely, Los Angeles, 
London, New York City and Paris.
The size of the grid is comparable to what is reported in the pRide paper.
We have done experiments with the number of drivers as 5, 15 and 25
per grid, in each case distributed randomly throughout each grid
but located on motorable areas.
We note here that the number of drivers does not have a bearing on our attack
since the SP encrypts and blinds each driver's distances independent
of one other.

In each of the maps, we picked random driver locations situated on motorable roads.
Next, a rider location was picked from a random grid in the map.  As explained
in Section \ref{tvt:retr_dist}, grids adjacent to the rider's grid was 
examined and distances between drivers in those grids and the rider were made
available to the rider.  Except for the predicted result ($PR$) values, this is 
same as what is available to the rider in the pRide protocol.
The $PR$ values do not have any bearing on our attack 
since they do not have any effect on either blinding or encryption of distances.

Next, from each of the adjacent grids and for each driver in such grid, the
distances from each such driver to her respective grid corners were computed,
and blinded using random integers picked from the range $[1, 2^{24}]$, as
the maximum UTM (northing) value of $10^{7}$ can be represented using 24 bits.
In addition, a distance value known to the adversary is also blinded using
the same random integers. 
These blinded distances were made available to the adversary rider.  Again,
this exactly mimics the behaviour of the enhanced pRide protocol.

Finally, we run the attack described in Section \ref{sec:our_attack} to
retrieve the distances followed by retrieving the driver locations.\\
\textit{Remark 1}:  It is claimed that the security of the pRide protocol relies on the hardness
of obtaining the blinding parameters when given only the blinded values.
We show in our attack that the adversary can recover the blinding parameters
with a high probability.   \\
\textit{Remark 2}: In our experiments, we have used a search radius
$SR =1$.  Our attack methodology can be easily extended to higher values of search radius.
Since the order of grid traversal is known a priori, the new attack has to
compute equiprobable locations in each of the possible grids and continue with
our driver retrieval attack, as described in Section \ref{tvt:retr_locs}.

\subsection{Results}
The results of our experiments are tabulated in Table \ref{pride:tab:results}.
The pRide paper uses a $64\times 64$
grid over the city of Chengdu, China, and mentions a maximum of 16000 drivers
in their experiments, which translates to about 4 drivers per grid on average.
As it can be much larger in high density areas in the city,
we run our experiments with 5, 15 and 25 drivers per grid.  
It takes less than 1 second to recover the locations of 25 drivers.

In order to retrieve the distances, we first recover the blinding integers
$e$ and $r$ as described in Section \ref{tvt:retr_dist}.  As shown in
Table \ref{pride:tab:results}, we can retrieve at least 80\% of the distances 
successfully, averaged from 10 runs of the experiments for each driver count over each city.
In the unsuccessful cases, we find that the value of the
blinding value $e$
retrieved by our algorithm is a multiple of the actual value of $e$,
and we report this as a failure.

Next, we use the successfully retrieved distances to obtain the
precise driver locations.  Here, we use our attack described in
Section \ref{tvt:retr_locs}.  We see that this part correctly retrieves close to 99\%
of the driver locations.
Hence, our overall driver location harvesting algorithm retrieves at least
80\% of the drivers participating in the enhanced pRide protocol.

\begin{table}[htpb]
\caption{Percentage of driver locations recovered for multiple cities.}
\label{pride:tab:results}
\begin{center}
\begin{tabular}{l c c}
\hline
City & Number of participating & \%age of driver\\
& drivers (per grid)  & coordinates correctly recovered \\
\hline
Los Angeles & 5 & 80\\
& 15 & 95 \\
	    & 25 & 89 \\
\hline
London &     5 & 85\\
& 15& 81\\
	    & 25& 86 \\
\hline
New York City &     5 & 90\\
& 15& 95 \\
	    & 25& 93\\
\hline
Paris &     5 & 85\\
& 15& 93\\
	    & 25& 88\\
 \hline
\end{tabular}
\end{center}
\end{table}

\section{Related Works}
\label{sec:rel_works}
There is a large body of work on privacy-preserving RHS which consider preserving
privacy of drivers and riders.
ORide \cite{ORidePaper} and PrivateRide \cite{Pham2017PrivateRideAP}, both
proposed by Pham \etal\hspace{-0.05in}, were some of the early works that aimed to preserve
rider privacy against SP and drivers.  While PrivateRide makes use of a
cloaking region to maintain privacy, ORide scheme is based on SHE to encrypt
driver and rider locations so as to make use of homomorphic properties of SHE to
select nearest driver.
Kumaraswamy \etal\cite{deepakOridePaper} proposed an attack that 
aims to determine locations of drivers participating in the ORide protocol.
In their attack, an adversary rider can reveal locations of up to 40\% 
of drivers who respond to a single ride request.  
They provide a countermeasure to thwart the attack while preserving 
sufficient anonymity.  Murthy \etal\cite{murthyOrideAttack} proposed an
attack that uses triangulation by four colluding adversaries to obtain locations
of all drivers participating in the ORide protocol.
%The basic pRide protocol \cite{pRideHuang2021}, described in detail
%in Section \ref{sec:pride_prot}, also makes
%use of a method similar to ORide to report rider to driver distances
%to the rider.  We feel the attack in \cite{deepakOridePaper} might also
%be applicable to the basic pRide protocol.

Luo \etal \cite{pRideLuo} proposed a privacy-preserving ride-matching service 
also named \textit{pRide}.  Their protocol involves using two non-colluding 
servers: SP and CP (a third-party crypto server), and uses 
Road Network Embedding (RNE) \cite{shahabiRNE} such that the road network
is transformed to a higher dimension space to enable efficient distance computation between the
network entities.
However, the disadvantage of their scheme is the use of two non-colluding
servers which incurs inter-server communication costs.
Yu \etal \cite{lpRideYu} proposed lpRide protocol which also uses RNE but
uses a modified version of Paillier encryption scheme \cite{nabeelPaillier}
to preserve privacy of participating entities.
Vivek \cite{svivek_attack} demonstrated an attack on the lpRide protocol
where they show that any rider or driver can learn the coordinates of
other participating riders.
TRACE \cite{wangTrace} is a privacy-preserving dynamic spatial query 
RHS scheme proposed by Wang \etal\hspace{-0.05in}, that uses a quadtree structure and
provides high-efficiency in terms of complexity and communication overhead.
Kumaraswamy \etal \cite{deepakIndocrypt} demonstrated an attack on the
TRACE protocol where the SP can identify the exact locations of riders and drivers.
Xie \etal \cite{xieTifs2021} proposed a protocol that also uses RNE to 
efficiently compute shortest distances.  Their scheme makes use of
property-preserving hash functions where the SP can not only compute 
the rider to driver distances, but also pick the nearest driver. 
This way they eliminate the need for an auxiliary crypto server.
All the works listed earlier picks the nearest driver to fulfil a
ride request.  pRide \cite{pRideHuang2021}, proposed by Huang \etal\hspace{-0.05in},
does not match the nearest driver but considers a global matching strategy 
with the aim of reducing the empty distance travelled by driver to pick the rider.
Murthy \etal \cite{murthyOrideAttack} gave an attack on the ORide protocol, using
triangulation, where they recover locations of all participating drivers.  In addition,
by using more number of colluding adversaries, they show they can recover locations
of up to 50\% of drivers participating in the variant of ORide protocol that uses
the mitigation solution of \cite{deepakOridePaper}.

\section{Conclusions}
\label{sec:conc}
In this paper, we presented an attack on enhanced pRide \cite{pRideHuang2021} protocol, 
a privacy-preserving RHS.  We show that an honest-but-curious adversary rider
can determine the coordinates of about 80\% of drivers responding to
the rider's ride request as per the pRide protocol.
%Our attack involves retrieving the blinding parameters followed by
%recovering the distances.  Finally, using the distances, the precise locations
%of the drivers are recovered by using the rider to driver distance
%and Google Nearest Road APIs to eliminate locations that are not on motorable
%areas.

From Section \ref{tvt_attack:oride_ref}, we see that locations of all drivers
participating in the basic pRide protocol can be recovered by one or more
adversary riders.  As per the protocol, the rider chooses the optimum driver
when given the plaintext distances to all drivers, and this fact is exploited by the adversary.
Alternatively, the SP can select the optimum driver homomorphically.  Since
sorting and searching are high-depth circuits, it is not efficient to perform
these operations using SHE schemes.  However, FHE schemes can be 
explored to evaluate their suitability for practical RHS solutions.

The enhanced pRide protocol needs to perform comparisons and in order to
preserve privacy, the values are blinded.  However, since the order needs to be
preserved, the blinding values are the same for all the comparands, which
leads to the attack.  Other secure order-preserving techniques need to be explored.
However, as shown in \cite{MurthyV19}, careful analysis is needed which would
otherwise lead to further attacks.

In summary, we show that although protocols may seem secure in theory,
a thorough analysis should be done which otherwise would expose severe
vulnerabilities and security holes, as demonstrated by our attack
in this paper.

\subsubsection*{Acknowledgements.}
We thank the anonymous reviewers for their review comments.
This work was partly funded by the Infosys Foundation Career Development Chair Professorship grant for Srinivas Vivek.

\bibliographystyle{splncs04}
\bibliography{abbrev0,crypto,morerefs}

%Author, F.: Article title. Journal \textbf{2}(5), 99--110 (2016)

\end{document}